\documentstyle[aps,epsfig]{revtex}
\newcommand{\rr}{\vec{r}_0}
\newcommand{\rone}{\vec{r}_1}
\newcommand{\ueff}{U_{\mathrm{eff}}}
\newcommand{\vp}{\varphi}
\renewcommand{\ao}{a_{\mathrm{osc}}}

\begin{document}
\draft
\preprint{ }

\title{Hydrodynamic Approach to Vortex Lifetime in Trapped Bose Condensates}
\author{Emil Lundh and P.\ Ao}
\address{Department of Theoretical Physics,  Ume{\aa} University,
S-90187 Ume{\aa}, Sweden
\\}

\date{\today}

\maketitle

\begin{abstract}
We study a vortex in a two-dimensional, harmonically trapped 
Bose-Einstein condensate at zero temperature. Through a variational 
calculation using a trial condensate wave function and a
nonlinear Schr\"odinger Lagrangian, we obtain the 
effective potential experienced by a vortex at an arbitrary position 
in the condensate, and find that an off-center vortex will move in
a circular trajectory around the trap center. We find the frequency 
of this precession to be
smaller than the elementary excitation frequencies in 
the cloud.

We also study the radiation of sound from a moving vortex in an
infinite, uniform system, and discuss the validity of this
as an approximation for the trapped case. Furthermore, we estimate
the lifetime of a vortex due to imperfections in the trapping 
potential.

   \end{abstract}


\section{Introduction}
The prospect of creating quantized vortices in trapped 
Bose-Einstein condensed gases (BEC's) has been an intensely discussed and 
studied subject in the last few years
\cite{italianrmp,rokhsar,mzh,fetter,pu,svidzinsky,feder}.
Despite the considerable interest in this area of BEC studies, many of the
most fundamental questions have yet to be answered, such as those 
concerning the stability and lifetime of such a state 
\cite{rokhsar,fetter,pu,svidzinsky,feder}.
In this article, we study the properties of a vortex in a trapped
BEC from a hydrodynamic point of view. We confine the discussion 
to two-dimensional systems at zero temperature,
and furthermore employ the limit of a condensate which is
large in comparison 
with the size of the vortex.

We will thus only be concerned with a system whose properties can
be described by a nonlinear Schr\"odinger equation. The system 
that we have in mind is the dilute Bose gas, which is governed by
the Gross-Pitaevskii equation \cite{gross,pitaevskii} 
at temperatures 
sufficiently low that
the system may be described as a superfluid. 
It should, however, be noted
that this type of equation
is applicable to a wider class of systems than just zero-temperature
dilute gases \cite{demircan,aitchison}.

The stability of a vortex in a BEC is limited by several factors.
At finite temperatures, the vortex 
may be destroyed due to collisions with thermal
excitations. \cite{fedichev} 
This will not be the subject of this study. Second, the
vortex may decay spontaneously even at zero temperature through
the excitation of modes (or, equivalently, the emission of phonons)
in the cloud \cite{rokhsar,feder}. Third, deviations from spherical symmetry
in the trapping potential will also limit the lifetime of the vortex. The
two latter processes will be the subject
of this paper.

The paper is organized as follows.
Sections \ref{lagrangesec}-\ref{lifetimesec} are concerned
with the motion of an off-center vortex and its decay through
phonon emission.
In section \ref{lagrangesec} we introduce the model and the
trial assumptions for the density and velocity distributions, 
and analyze the motion of an off-center vortex. In 
Sec.\ \ref{radiationsec}, the parallel case of a precessing vortex
in an infinite, homogeneous system is analyzed, and the power loss
due to the radiation of sound is calculated. This can be thought of
as a ``semiclassical'' approximation to the trapped case considered 
here. In Sec.\ \ref{lifetimesec}, the validity of this approximation
is discussed and a lower bound for the lifetime of a vortex
is arrived at. In Sec.\ \ref{lsec}, we find the 
characteristic time for destruction
of a vortex due to deviations from cylindrical symmetry in the trapping
potential, and finally, in Section \ref{conclusionsec}, the results are 
summarized and discussed.

\section{Circular motion}
\label{lagrangesec}

In the case of a harmonic trapping potential, the 
equation for the condensate wave
function $\psi(\vec{r})$, whose squared modulus gives the
superfluid density
distribution $\rho(\vec{r})$, reads
  \begin{eqnarray}
 \left[
   -\frac{\hbar^2}{2m} \nabla^2
   + \frac12 m \omega_t^2 r^2
    +U_0|\psi(\vec{r})|^2\right]\psi(\vec{r}) = \mu \psi(\vec{r}),
\label{gp}
\end{eqnarray}
where $\mu$ is the chemical potential and $U_0 = 4 \pi \hbar^2 a/m$
is the effective interaction potential, with $a$ being the 
$s$-wave scattering length. 
Assuming the kinetic energy to be negligible, we obtain the so-called
Thomas-Fermi approximation \cite{bp} for the wave function 
for a non-rotating cloud,
\begin{equation}
  \label{tfwf}
  \psi_{\mathrm{TF}}(\vec{r}) = \sqrt{\rho_0} 
  \left(1-\frac{r^2}{R^2}\right)^{1/2},
\end{equation}
with an associated density distribution 
$\rho_{\mathrm{TF}} = |\psi_{\mathrm{TF}}|^2$.
Here, the central density $\rho_0 = \mu/U_0$ and the Thomas-Fermi radius 
$R = \left(2 \mu/m \omega_t^2\right)^{1/2}$.
For a two-dimensional system, the wave function $\psi(\vec{r})$ is
normalized according to
\begin{equation}
  \label{normalization}
  \int d^2 \vec{r} \left| \psi(\vec{r})\right|^2 = \nu,
\end{equation}
where $\nu$ is the number of particles per unit length. As a 
measure of the influence of the inter-particle interactions
on the system's properties, we define the 
dimensionless parameter
\begin{displaymath}
  \gamma \equiv \nu a.
\end{displaymath}
The Thomas-Fermi approximation is a
valid one when  $\gamma$ is large, and in that limit we have
for a two-dimensional system
\begin{eqnarray}
\nonumber 
\mu = 2 \hbar \omega_t \sqrt{\gamma},\\
\label{tfquantities}
R = \ao 2 \gamma^{1/4}, 
\end{eqnarray}
where $\ao= \left(\hbar / m \omega_t \right)^{1/2}$ is the oscillator 
length.

We now turn to the problem of
a cloud containing a singly quantized vortex at the position $\rr$.
In the limit of large $\gamma$, the density distribution of the cloud 
will not be appreciably affected by the presence of a vortex, except
in a region whose size is comparable to the healing length $\xi$, 
defined as
\begin{equation}
  \label{healinglength}
  \xi = \sqrt{\frac{\hbar^2}{2m\rho U_0}} =\frac1{(8\pi \rho a)^{1/2}}.
\end{equation}
The healing length
gives the length scale over which the wave function for a vortex in
a homogeneous Bose gas increases from zero to its bulk 
value \cite{gross,pitaevskii,ginzburgp}. 
For an untrapped system, the density $\rho$ is the value of
the density far from the vortex core. In the case of a trapped
system, $\rho$
must be taken to be the local Thomas-Fermi density
at the point $\rr$ in the absence of a vortex, $\rho(\rr)$, thus
defining a local healing length
\begin{equation}
  \label{localcoherencelength}
  \xi(r_0) = \frac1{(8\pi \rho(r_0) a)^{1/2}} = 
  \frac{\xi_0}{\sqrt{1-\frac{r_0^2}{R^2}}},
\end{equation}
where $\xi_0 = \xi(0)$ is the value of the healing length in
the center.
The velocity distribution in a Bose-condensed system is given by
$\hbar/m$ times the gradient of the phase of the wave function. 
For a positively oriented vortex in an infinite, 
uniform system, it is known to be
\begin{displaymath}
\vec{v}_{\mathrm{uni}}(\vec{r}) = \frac{\hbar}{m} \nabla \phi,
\end{displaymath}
where $\phi$ is the polar angle relative to the position of the vortex,
which gives
\begin{displaymath}
\vec{v}_{\mathrm{uni}}(\vec{r}) = \frac{\hbar}{m}
\frac{\hat{z} \times (\vec{r}-\rr)}{|\vec{r}-\rr|^2},
\end{displaymath}
$\hat{z}$ being the unit vector in the $z$ direction.

The velocity field is altered due to the boundary of the system
and due to the spatially varying density. The existence of a boundary
requires that the normal velocity vanishes there, and for homogeneous
systems the recipe is to introduce a negatively oriented image 
vortex at the point
$\rone \equiv \rr R^2/r_0^2$, giving the velocity field
\footnote{The existence of sharp boundary is an artefact of the Thomas-Fermi
approximation. Although this approximation never holds at the
boundary, we do have that for sufficiently large clouds, $\rho(R)$ 
is small, while $\nabla \rho(R)$ is nonnegligible, which justifies the
neglect of the first term in the equation for stationary flow, 
$\rho \nabla\cdot \vec{v}+\vec{v}\cdot\nabla \rho=0$. Hence 
the radial velocity has to (approximately) vanish at $r=R$.}
\begin{equation}
\vec{v}_0(\vec{r}) = \frac{\hbar}{m}
             \frac{\hat{z} \times (\vec{r}-\rr)}{|\vec{r}-\rr|^2}
                 - \frac{\hbar}{m}
             \frac{\hat{z} \times (\vec{r}-\rone)}{|\vec{r}-\rone|^2}.
\label{velocity0}
\end{equation}
If the system has a density gradient, as in the present case, 
the condition for stationary flow, $\nabla \cdot (\rho \vec{v})=0$, is not 
automatically fulfilled. Writing the velocity field as
\begin{displaymath}
  \vec{v} = \vec{v}_0 + \vec{v}_1,
\end{displaymath}
with $\vec{v}_0$ taken from Eq.\ (\ref{velocity0}),
we get an equation for the correction $\vec{v}_1$:
\begin{equation}
\label{veleq1}
  \rho \nabla\cdot\vec{v}_1 + \vec{v}_0 \cdot\nabla\rho + 
  \vec{v}_1 \cdot\nabla\rho = 0,
\end{equation}
since the divergence of $\vec{v}_0$ vanishes.
An approximate solution to this equation, valid close to the center
of the system, may be found by treating $\nabla\rho$ as small,
whereupon $\vec{v}_1$ can also be expected to be a small correction,
and the third term on the left-hand side of Eq.\ (\ref{veleq1})
can be discarded. We then have 
\begin{displaymath}
  \nabla \cdot \vec{v}_1(\vec{r}) = f(\vec{r}),
\end{displaymath}
where 
\begin{displaymath}    
  f(\vec{r}) =
  - \frac{\vec{v}_0(\vec{r}) \cdot\nabla\rho(\vec{r})}{\rho(\vec{r})},
\end{displaymath}
with the boundary condition that the normal velocity vanish for $r=R$.
Since we consider a Bose condensate, the velocity has to be a 
potential flow, $\vec{v}_1(\vec{r}) = \nabla \phi_1(\vec{r})$, and we have
the equation for $\phi_1$,
\begin{eqnarray*}
  \nabla^2 \phi_1(\vec{r}) = f(\vec{r}) \mbox{ for } r \le R,\\
  \frac{\partial \phi_1(\vec{r})}{\partial r} = 0 \mbox{ for } r= R
\end{eqnarray*}
whose solution is written in terms of the Green's function for the 
Neumann problem on a disk of radius $R$
\begin{displaymath}
  G_N(\vec{r}', \vec{r}) = 
  \frac1{2\pi}\ln \frac{R}{|\vec{r}'-\vec{r}|}
  + \frac1{2\pi}\ln \frac{R}{\left|\vec{r}'-\frac{R^2}{r^2}\vec{r}\right|}
  + \frac1{2\pi}\ln \frac{R}{r}
\end{displaymath}
as
\begin{equation}
  \label{phiprime}
  \phi_1(\vec{r}) = \int d^2 \vec{r}' f(\vec{r}') G_N(\vec{r}', \vec{r}).
\end{equation}
One can easily obtain higher-order velocity terms in $\nabla \rho$,
if one writes
\begin{displaymath}
  \vec{v} = \vec{v}_0+\vec{v}_1+\vec{v}_2+...
\end{displaymath}
One immediately finds that
\begin{displaymath}
  \phi_{n+1}(\vec{r}) = -\int d^2 \vec{r}' 
  \frac{\vec{v}_n(\vec{r})\cdot\nabla\rho(\vec{r})}{\rho(\vec{r})} 
  G_N(\vec{r}', \vec{r}),
\end{displaymath}
where $\vec{v}_n = \nabla \phi_n$, $n=1,2,...$ . 
When the density varies over a scale larger than the healing length,
higher order corrections are small.
We will be content in this paper to retain
only the zeroth-order term.

The energy per unit length of a two-dimensional system described by a 
 nonlinear Schr\"odinger equation is
\begin{eqnarray}
  \label{energy}
  E[\psi, \rr] = \int d^2\vec{r} \left( 
    \frac{\hbar^2}{2m}\left|\nabla\psi\right|^2
    + \frac12 m \omega_t^2 r^2 \left|\psi\right|^2
    + \frac{U_0}{2} \left|\psi\right|^4
  \right),
\end{eqnarray}
where we have made explicit the dependence of $E$ on the vortex
coordinate $\rr$.
The change in energy of the system due to the presence of a vortex
only shows up in the kinetic-energy term, as long as its effect 
on the density profile is neglected. We shall
denote this additional energy by $\ueff$, since one may regard it
as an effective potential for the vortex, depending on the vortex
position $\rr$. In terms of the velocity field $\vec{v}(\vec{r})$
it is written
\begin{equation}
  \ueff(\rr) = \frac{m}{2} \int d^2\vec{r} \rho(\vec{r}) v(\vec{r})^2.
\end{equation}
Using the lowest-order approximation $\vec{v}_0$ for the velocity,
and employing the Thomas-Fermi wave function (\ref{tfwf}),
the result is (cf. \cite{fetter})
\begin{eqnarray}
  \ueff(\rr) &=& \frac{\pi \hbar^2 \rho_0}{2m} \left[
  \left(1-\frac{r_0^2}{R^2}\right)\ln\left(\frac{R^2}{\xi_0^2}\right)
    + \left(\frac{R^2}{r_0^2}+1-2\frac{r_0^2}{R^2}\right)
    \ln\left(1-\frac{r_0^2}{R^2}\right)
  \right].
  \label{ueff}
\end{eqnarray}
To obtain this result, one needs to exclude the vortex core,
of size $\xi(r_0)$ around the vortex position, 
from the radial integral for the integrals to converge.
The first term dominates for small $r_0$ and is independent of the
details of the model.

The above analysis shows that as long as the system is not subjected 
to an external rotational constraint,
the vortex will experience an effective potential
which decreases with the distance from the trap center, except in
a small region close to the boundary, where $\ueff$ has a
local minimum due to the unphysical behaviour of the Thomas-Fermi
wave function at $r=R$. This minimum is located a distance
$\delta$ from the edge, given by 
\begin{equation}
\label{delta}
  \delta = \frac{1}{2e}\left(\frac{\xi_0}{R}\right)^{2/3}.
\end{equation}
It is interesting to note that the healing length does not set
the length scale here. Apart from constants of order unity,
this ``boundary thickness'' is the same as the cut-off length
$\delta$ of Ref.\ \cite{lps}, which comes about when computing 
the kinetic energy of a Thomas-Fermi cloud.

We now turn to the motion of the vortex in this simple model.
We assume that the vortex coordinate $\rr$ 
may have a time dependence, but that all other parameters of the
system remain stationary.
Solving the time-dependent counterpart to the equation 
(\ref{gp}) is equivalent to minimizing the action obtained from the
Lagrangian \cite{demircan,perez}
\begin{equation}
  \label{lagrangian}
  L[\psi,\rr,\dot{\rr}] = T[\psi,\rr,\dot{\rr}] - E[\psi,\rr],
\end{equation}
where the kinetic term
\begin{equation}
  \label{kdef}
  T[\psi,\rr,\dot{\rr}] = \int d^2\vec{r}\,
  \frac{i\hbar}{2} \left[ \psi^* \frac{\partial \psi}{\partial t}
  - \psi \frac{\partial \psi^*}{\partial t}\right].
\end{equation}
A straightforward calculation, remembering that 
the gradient of the phase of $\psi$ is the velocity field, yields
\begin{equation}
  T = \hbar\frac{\hat{z}\cdot(\dot{\rr}\times\rr)}{r_0^2}
  \left(\tilde{\nu}(\rr)-\nu\right),
\end{equation}
where $\tilde{\nu}(\rr)=2\pi\int_0^{r_0}dr\,r \rho(r)$ is the number
of particles
per unit length inside the circle of radius $r_0$.
The Euler-Lagrange 
equation for $\rr$ and $\dot{\rr}$ will finally yield,
for the radial ($r_0$) and azimuthal ($\phi_0$) components respectively,
\begin{eqnarray*}
\dot{r}_0 = 0;\\
  \dot{\phi}_0 =  \frac{F(r_0)}{\hbar \partial\tilde{\nu}/\partial r_0}
\end{eqnarray*}
where we have defined $F= -\partial \ueff/\partial r_0$. We see that
in this model, an off-center vortex executes an orbiting motion around the
center with an angular frequency $\omega = \dot{\phi}_0$. Since 
$\partial\tilde{\nu}/\partial r_0 = 2\pi r_0\rho(r_0)$, we finally
obtain
\begin{displaymath}
  \omega = \frac{F(r_0)}{2\pi\hbar r_0 \rho(r_0)}.
\end{displaymath}

In the case of a Thomas-Fermi profile, $F$ is obtained by
differentiating Eq.\ (\ref{ueff}):
\begin{equation}
  \label{f}
  F(r_0) = \frac{\pi\hbar^2 \rho_0 r_0}{m R^2} g(r_0/R),\\
\end{equation}
where
\begin{equation}
  \label{g}
  g(x) = 2 \ln\left(\frac{R}{\xi}\right) + 
  \left(\frac{1}{x^4}+2\right) \ln\left(1-x^2\right)+\frac{1}{x^2}+2,
\end{equation}
which yields the final result for the frequency of precession
of a vortex in a Thomas-Fermi cloud,
%
\begin{equation}
  \label{frequency}
  \omega = \frac{\hbar}{2 m R^2 \left(1-\frac{r_0^2}{R^2}\right)}g(r_0/R).
\end{equation}
The same expression has been obtained previously by different approaches 
\ \cite{rp,svidz2}. Nevertheless, 
the above treatment allows a smooth generalization to 
incorporate the phonon effect.

\section{Phonon radiation by a vortex in an infinite system}
\label{radiationsec}

We now turn to the problem of a vortex in an infinite system,
exercising (e.\ g.\ under the influence of an external force)
circular motion.

It has previously been shown \cite{arovas,ambegaokar} 
how any homogeneous superfluid described by 
a nonlinear  Schr\"odinger-type energy functional is equivalent to 
(2+1)-dimensional electrodynamics, with vortices playing the 
role of charges and sound corresponding to electromagnetic radiation. 
For a fluid, which in the
absence of vortices has the density $\rho_0$ (note
that the use of this symbol is not the same as in the preceding 
section), with the local fluid
velocity $\vec{v}(\vec{r},t)$ and density $\rho(\vec{r},t)$, and 
(possibly) containing a vortex at the position $\rr(t)$, moving at a
velocity $\dot{\rr}(t)=\vec{v}_v(t)$, we define the ``vortex charge'' 
$q_v = -\hbar\sqrt{2\pi \rho_0/m}$, ``vortex 
density'' $\rho_v(\vec{r},t) = \delta^{(2)}(\vec{r}-\rr(t))$, 
and the corresponding ``vortex current'' $\vec{\jmath}_v(\vec{r},t) = 
q_v \rho_v(\vec{r},t) \vec{v}_v(\vec{r},t)$. The speed of sound
is $c=\sqrt{U_0\rho_0/m}$.
We then have the analogous Maxwell equations:
\begin{eqnarray*}
  \nabla \cdot \vec{b} &=& 0,\\
  \nabla \cdot \vec{e} &=& 2\pi q_v \rho_v,\\
  \nabla \times \vec{e} + \frac{1}{c} \frac{\partial \vec{b}}{\partial t}
  &=& 0,\\
  \nabla \times \vec{b} - \frac{1}{c} \frac{\partial \vec{e}}{\partial t}
  &=& \frac{2\pi}{c} q_v \vec{\jmath}_v,
\end{eqnarray*}
where we have defined
\begin{eqnarray*}
  \vec{e}(\vec{r},t) &=& \sqrt{\frac{2\pi m}{\rho_0}} \rho(\vec{r},t)
  \hat{z}\times \vec{v}(\vec{r},t),\\
  \vec{b}(\vec{r},t) &=& \sqrt{\frac{2\pi m}{\rho_0}} c 
  \hat{z} \rho(\vec{r},t).\\
\end{eqnarray*}
The ``no magnetic monopole'' law is clear from the definition of
$\vec{b}$; the Coulomb law states how the presence of vortices
create a rotational current, the Faraday law is equivalent to
the continuity equation for the fluid, and the counterpart to
Amp{\`e}re's law derives from the Josephson-Anderson relation 
implied by the Euler equation.
The energy of the system is
\begin{displaymath}
  E = \frac{1}{4\pi} \int d^2\vec{r} \left(\vec{e}^2(\vec{r},t) +
  \vec{b}^2(\vec{r},t)\right),
\end{displaymath}
and correspondingly the Poynting vector
\begin{equation}
  \label{poynting}
  \vec{\sigma}(\vec{r},t)= \frac{c}{2\pi}\vec{e}(\vec{r},t)
  \times\vec{b}(\vec{r},t).
\end{equation}

Electromagnetic potentials, $\vec{a}(\vec{r},t)$ and $\vp(\vec{r},t)$,
are defined in the usual way, and within a Lorentz gauge we
recover the usual wave equations
\begin{eqnarray*}
  \left(\nabla^2 -\frac{1}{c}\frac{\partial^2}{\partial t^2}\right) 
    \vp(\vec{r},t) &=& -2 \pi \rho_v(\vec{r},t),\\
  \left(\nabla^2 -\frac{1}{c}\frac{\partial^2}{\partial t^2}\right) 
    \vec{a}(\vec{r},t) &=& -\frac{2 \pi}{c} \vec{\jmath}_v(\vec{r},t),\\
\end{eqnarray*}
which have the exact solution in (2+1) dimensions \cite{morse}
\begin{eqnarray}
\nonumber
  \vp(\vec{r},t) &=& -\int dt' d^2\vec{r}'
  \frac{\theta\left(t-t'-\frac{|\vec{r}-\vec{r}'|}{c}\right)}{
    \sqrt{(t-t')^2-\frac{|\vec{r}-\vec{r}'|^2}{c^2}}}\rho_v(\vec{r}',t'),\\
\label{exact2d}
\vec{a}(\vec{r},t) &=& -\frac{1}{c}\int dt' d^2\vec{r}'
  \frac{\theta\left(t-t'-\frac{|\vec{r}-\vec{r}'|}{c}\right)}{
    \sqrt{(t-t')^2-\frac{|\vec{r}-\vec{r}'|^2}{c^2}}}
  \vec{\jmath}_v(\vec{r}',t').
\end{eqnarray}
The step functions $theta$ in the numerator is a feature peculiar to two
dimensions; the corresponding three-dimensional expressions contain
a delta function.

We wish to use the above formulation of vortex dynamics in order to
find the energy dissipated from a circularly moving vortex due to
the radiation of sound waves. Our aim is therefore to find the value
of the Poynting
vector $\vec{\sigma}$ at large distances from the vortex.

We first specialize to the case of one point particle at the position 
$\rr$. Eq.\ (\ref{exact2d}) becomes
\begin{eqnarray}
\nonumber
  \vp(\vec{r},t) &=& - q_v\int dt'
  \frac{\theta\left(t-t'-\frac{X}{c}\right)}{
    \sqrt{(t-t')^2-\frac{X^2}{c^2}}},\\
  \label{integrals2d}
\vec{a}(\vec{r},t) &=& -\frac{q_v}{c}\int dt'
  \frac{\theta\left(t-t'-\frac{X}{c}\right)}{
    \sqrt{(t-t')^2-\frac{X^2}{c^2}}}
  \vec{v}_v(t'),
\end{eqnarray}
where $X = |\vec{r}-\rr(t')|$. We now seek the solutions to 
these integrals in the limit of large $\vec{r}$. To zeroth order
in $r_0/r$, $X=r$, which is independent of $t'$.
For a charge exercising circular motion with frequency $\omega$ at
a radius $r_0$, the velocity is
\begin{displaymath}
  \vec{v}_v(t') = v(-\hat{x}\sin\omega t' + \hat{y}\cos\omega t').
\end{displaymath}
The integrals for the vector potential's $x$ and $y$ components can
now be done exactly, yielding Bessel functions:
\begin{eqnarray*}
a_x(\vec{r},t) = \frac{q_v v}{c}\int_{-\infty}^{t-r/c} 
\frac{dt' \sin\omega t'}{\sqrt{(t-t')^2-\frac{r^2}{c^2}}}
= -\frac{qv\pi}{2c}\left[N_0\left(\frac{\omega r}{c}\right)\sin \omega t
+ J_0\left(\frac{\omega r}{c}\right)\cos \omega t\right],\\
a_y(\vec{r},t) = -\frac{q_v v}{c}\int_{-\infty}^{t-r/c} 
\frac{dt' \cos\omega t'}{\sqrt{(t-t')^2-\frac{r^2}{c^2}}}
= -\frac{qv\pi}{2c}\left[-N_0\left(\frac{\omega r}{c}\right)\cos \omega t
+ J_0\left(\frac{\omega r}{c}\right)\sin \omega t\right],
\end{eqnarray*}
and finally
\begin{displaymath}
  \vec{a}(\vec{r},t) = \frac{\pi q_v}{2c} \left[\vec{v}_v(t)
  N_0\left(\frac{\omega r}{c}\right) + \hat{z}\times\vec{v}_v(t)
  J_0\left(\frac{\omega r}{c}\right)\right].
\end{displaymath}
We immediately obtain the ``magnetic field'':
\begin{equation}
\label{bfield}
  \vec{b}(\vec{r}) = \frac{\pi q_v\omega}{2c^2}\left[\vec{v}_v(t)
  N_1\left(\frac{\omega r}{c}\right) +\hat{z}\times\vec{v}_v(t)
  J_1\left(\frac{\omega r}{c}\right)\right]\times \hat{r}.
\end{equation}
Finally, we utilize the asymptotic formulas for the Bessel functions
at large arguments, which yields
\begin{equation}
\label{asymptoticb}
  \vec{b}(\vec{r}) = q_v\sqrt{\frac{\pi \omega}{2c^3 r}}
  \left[\vec{v}_v(t)\times \hat{r}
    \sin\left(\frac{\omega r}{c}-\frac{3\pi}{4}\right) 
    +(\hat{z}\times\vec{v}_v(t))\times\hat{r}
    \cos\left(\frac{\omega r}{c}-\frac{3\pi}{4}\right)\right].
\end{equation}

It is not necessary to calculate the electric field $\vec{e}(\vec{r})$ 
in order to obtain the Poynting vector,
since at large $r$, the field
is locally that of a plane wave, in which case we have
\begin{equation}
  \label{poynting}
  \vec{\sigma} = \frac{c}{2\pi} \vec{b}^2 \hat{r}.
\end{equation}
On integrating around the circle with radius $r$ we get the power radiated
by a vortex exercising circular motion in an infinite system:
\begin{equation}
  \label{power}
  P = \int_0^{2\pi} r d\theta \hat{r} \cdot \vec{\sigma} =
  \frac{\pi q_v^2 \omega v^2}{4c^2} = \frac{\pi q_v^2 \omega^3 r_0^2}{4c^2}.
\end{equation}

\section{Lifetime of a vortex}
\label{lifetimesec}

In Sec.\ \ref{lagrangesec} we found that an off-center vortex performs
a circular motion. In the preceding section we saw how, in an infinite
and homogeneous system, such motion
excites sound waves, which carry away energy from the vortex. In a
trapped cloud, the effect of such radiation would be that the vortex 
move outward towards
regions of lower potential energy $\ueff(\rr)$, until it finally escapes
from the cloud \cite{rokhsar,feder}.

The application of the results of Sec.\ \ref{radiationsec} on
the trapped case may be thought of as a semiclassical approximation.
One condition for this approximation to hold is that the
precession frequency of the vortex match the attainable excitation 
frequencies of the cloud, as seen in Eq.\ (\ref{asymptoticb}), where
the moving vortex excites sound waves which have the same frequency 
as the precession. 

This requirement, however, is not met in the present case. 
In a harmonically trapped cloud containing a vortex, all but one of the mode 
frequencies are greater than or equal to the
trap frequency $\omega_t$ \cite{italianrmp,lecturenotes}; 
the single low-lying mode is identical to an
off-center displacement of the vortex \cite{fetter,svidzinsky,dodd}.

Comparing the precession frequency with the trap frequency, we find, 
using Eq.\ (\ref{tfquantities}),
\begin{displaymath}
  \frac{\omega}{\omega_t} = 
  \frac{g(r_0/R)}{8\gamma^{1/2}\left(1-\frac{r_0^2}{R^2}\right)}
\end{displaymath}
which is always less than one, except very close to the boundary.

We conclude that the semiclassical approximation is never valid for
a vortex in a harmonically trapped cloud\footnote{This is, 
in fact, the case for a larger class of trapping potentials, including all 
power-law potentials and the square well.}.
This does not, however, necessarily imply that the vortex is stable;
only that its lifetime is longer than that implied by the semiclassical
approximation.
 
The results of the two preceding sections can
therefore be utilized to calculate a lower bound for the vortex lifetime.
The power dissipated from the vortex
by phonon emission, Eq.\ (\ref{power}), is to be set equal to 
the rate of motion downhill the potential gradient:
\begin{displaymath}
  P = \frac{dE}{dt} = \frac{F dr_0}{dt}.
\end{displaymath}
We note that $P$ and $F$ are functions of $r_0$. Rearranging terms, we obtain
the time $\tau_p$ for the vortex to move from $r_0=\xi$ to $r_0=R-\delta$:
\begin{equation}
 \label{tau}
  \tau_p = \int_{\xi}^{R-\delta}dr_0 \frac{F(r_0)}{P(r_0)}.
\end{equation}
This quantity is a lower bound for 
the lifetime of a vortex originally residing
at a distance $\xi$ from the trap center. 
The upper cutoff, $R-\delta$,
is needed in order to avoid unphysical boundary effects, as 
discussed in connection with Eq.\ (\ref{delta}). The subscript
$p$ is introduced to indicate that this time scale is associated with
the radiation of phonons.

Inserting Eqs.\ (\ref{frequency}), (\ref{f}) and (\ref{power})
into (\ref{tau}) we get
\begin{displaymath}
  \tau_p = \frac{16 m^2 \rho_0 U_0 R^4}{\pi \hbar^3} I,
\end{displaymath}
where the dimensionless integral $I$ equals
\begin{displaymath}
  I = \int_{\xi/R}^{1-\delta/R} dx \frac{(1-x^2)^3}{x (g(x))^2},
\end{displaymath}
with $g(x)$ given by Eq.\ (\ref{g}). 
An exact result for the integral I is easily obtained numerically;
the result is shown in Fig. 1, and 
will be discussed shortly. An estimate
can be obtained by noting that the function $g(x)$
for strong coupling is approximately equal to $2 \ln( R/\xi)+\frac12$
over a large range of values of $x$, and that the lowest-order 
term in $x$ dominates the numerator, whereupon one gets
$I \approx \ln (R/\xi)/(2 \ln (R/\xi)+\frac12)^2$ and
\begin{equation}
\label{estimate1}
  \tau_p \approx \frac{16 m^2 \rho_0 U_0 R^4\ln (R/\xi)}
  {\pi \hbar^3 (2\ln(R/\xi)+\frac12)^2}.
\end{equation}
Finally, we insert the Thomas-Fermi results (\ref{tfquantities}),
to cast the above result in terms of the parameter $\gamma$:
\begin{equation}
\label{estimate}
  \tau_p \approx \frac{1}{\omega_t}\frac{128}{\pi}
  \frac{\gamma^{3/2}\ln (4\sqrt{\gamma})}{(\ln(4\sqrt{\gamma})+\frac14)^2}.
\end{equation}

\begin{figure}
\begin{center}
\epsfig{file=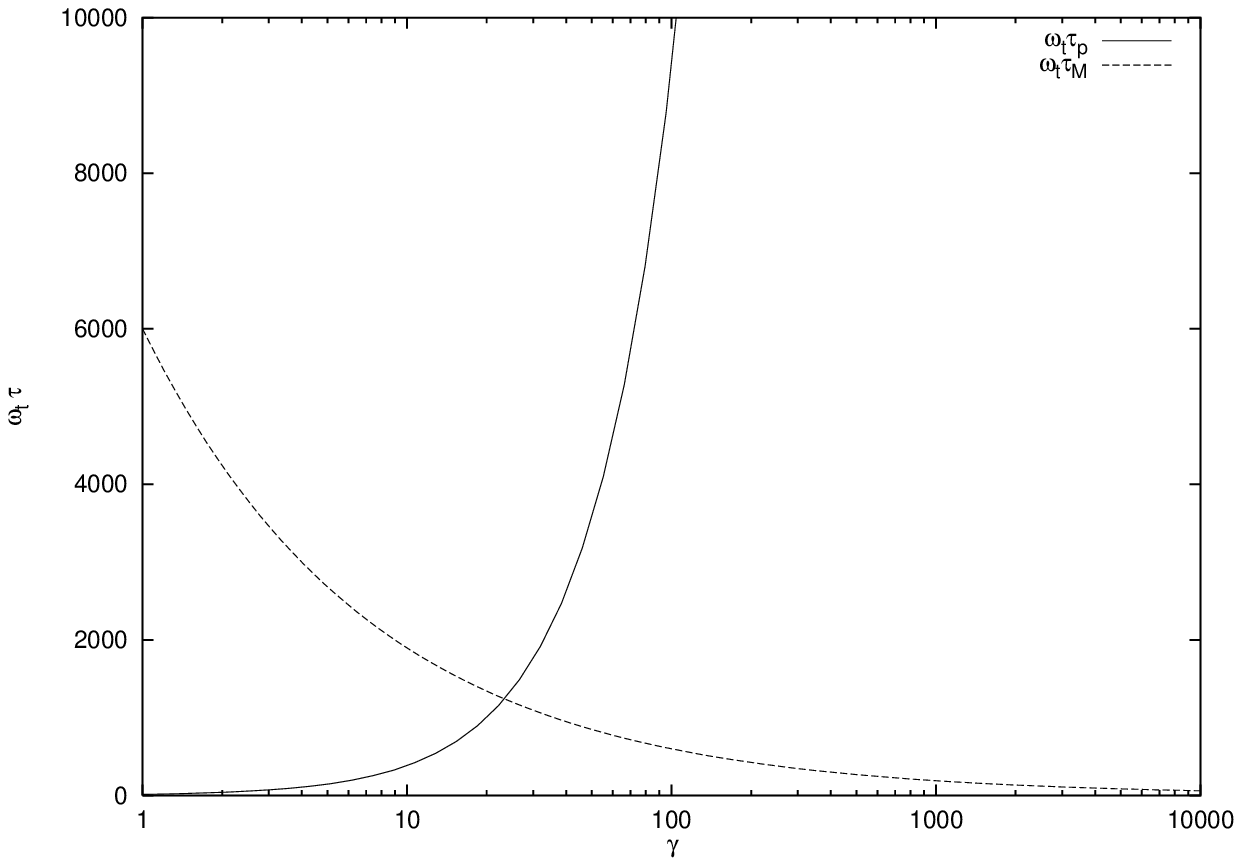,width=11cm,height=8cm,angle=0}
\begin{caption}
{Lower-bound estimates for the lifetime of the vortex, multiplied by
the trap frequency $\omega_t$. The solid
line shows the time scale $\tau_p$ connected with phonon radiation,
integrated numerically, while the dashed line is the quantity $\tau_M$
connected with broken rotational symmetry, with the parameter
$\epsilon $ set to 0.001.
}
\end{caption}
\end{center}
\label{FIG1}
\end{figure}

\section{ Broken rotational symmetry: change of angular momentum }
\label{lsec}

The other factor which at zero temperature may limit the lifetime
of a vortex is deviations from rotational symmetry of the trapping
potential. Even very small irregularities in the magnetic or electric
fields used to trap the condensed gases in experiments may affect
the possibility to keep a vortex in the system. 

We therefore
consider a BEC in a somewhat deformed trap. The total torque due to
inhomogeneities in the density and the trap is
\begin{equation}
  \label{torque}
  \vec{M} = \int d^2\vec{r} \rho(\vec{r}) \vec{r} \times \nabla V(\vec{r}),
\end{equation}
where $\rho(\vec{r})$ and $V(\vec{r})$ are the actual (not 
exactly cylindrically symmetric) density and external potential, respectively.
A natural unit for measuring the torque would be the potential
energy associated with the trapping potential, which has the same units:
\begin{displaymath}
  U_{\mathrm osc} = \int d^2\vec{r} \rho(\vec{r}) V(\vec{r}),
\end{displaymath}
whose value for a harmonic-oscillator trap and a
Thomas-Fermi density profile equals
\begin{displaymath}
 U_{\mathrm osc}  = \frac{\pi}{12} \rho_0 m\omega_t^2 R^4.
\end{displaymath}
Writing
\begin{displaymath}
  M = \vec{M}\cdot\hat{z} = \epsilon U_{\mathrm osc},
\end{displaymath}
we can use $\epsilon$ as a (approximately independent
of coupling strength) measure of the relative distortion of
the trap and the density profile. Assuming $\epsilon$ to be 
approximately constant over time, the time scale for the destruction
of an initially centrally placed vortex is
\begin{displaymath}
  \tau_M = \frac{L_0}{M},
\end{displaymath}
where $L_0$ is the value of the angular momentum for a system
with a central vortex; this is equal to $\hbar$ times the number of
particles $\nu$ per unit length; and so
the time for the vortex to move out is
\begin{displaymath}
  \tau_M = \frac{6 \hbar}{\epsilon m \omega_t^2 R^2}.
\end{displaymath}

\section{conclusions}
\label{conclusionsec}

We have arrived at two estimates for vortex lifetime: one connected
with phonon radiation, and one due to broken rotational symmetry.
The former, $\tau_p$, is 
an increasing
function of coupling strength $\gamma$, whereas the latter, 
$\tau_M$, decreases with increasing $\gamma$. This leaves us with a
``window'' at moderate values of coupling strength, where
both time scales are reasonably large.

The lower-bound time scales $\tau_p$ and $\tau_M$ are plotted
as functions of coupling strength in Fig.\ 1. 
Both are to be considered as lower bounds, since the assumptions
applied in deriving them are very pessimistic.
The parameter
$\epsilon$ is taken to be $10^{-3}$, which is an upper bound for the
trap deformations attainable experimentally \cite{wieman}. Trap 
frequencies are often around 100 s$^{-1}$ in experiments; thus $\tau_p$
will be longer than one second (wich is a typical order of magnitude
for condensate lifetimes) as long as $\gamma$ is greater than
unity, and $\tau_M$ is longer than one second for all $\gamma \le 5000$.
This leaves us with a large parameter range, easily attainable
experimentally, for which a vortex 
at zero temperature can
be considered as long-lived; considering the conservative 
assumptions made here, the actual region of vortex stability is 
probably much larger than this analysis indicates.

\section{Acknowledgement}
This work was in part supported by the Swedish NFR.
We would like to thank C.\ J.\ Pethick for
interesting discussions.

\end{document}